# Image reconstruction method for dual-isotope positron emission tomography

T. Fukuchi,[a,1] M. Shigeta,[a] H. Haba,[b] D. Mori,[b] T. Yokokita,[b] Y. Komori,[b] S. Yamamoto[c] and Y. Watanabe[a]

[a] *RIKEN Center for Biosystems Dynamics Research, Kobe 650-0047, Japan*
[b] *RIKEN Nishina Center for Accelerator-Based Science, Wako 351-0198, Japan*
[c] *Department of Radiological and Medical Laboratory Science, Nagoya University Graduate School of Medicine, Nagoya 461-0047, Japan*

   *E-mail*: tfukuchi@riken.jp

ABSTRACT: We developed a positron emission tomography (PET) system for multiple-isotope imaging. Our PET system, named multiple-isotope PET (MI-PET), can distinguish between different tracer nuclides using coincidence measurement of prompt γ-rays, which are emitted after positron emission. In MI-PET imaging with a pure positron emitter and prompt-γ emitter, because of the imperfectness of prompt γ-ray detection, an image for a pure positron emitter taken by MI-PET is superposed by a positron-γ emitter. Therefore, in order to make isolated images of the pure positron emitter, we developed image reconstruction methods based on data subtraction specific to MI-PET. We tested two methods, subtraction between reconstructed images and subtraction between sinogram data. In both methods, normalization for position dependence of the prompt γ-ray sensitivity is required in addition to detector sensitivity normalization. For these normalizations, we performed normalization scans using cylindrical phantoms of the positron-γ emitters $^{44m}$Sc (prompt γ-ray energy: 1157 keV) and $^{22}$Na (prompt γ-ray energy: 1274 keV). A long period measurement using the activity decay of $^{44m}$Sc ($T_{1/2}$ = 58.6 hours) elucidated that the acquisition ratio between the prompt γ-rays coincided with PET event and pure PET event changes on the basis of object activities. Therefore, we developed a correction method that involves subtraction parameters dependent on the activities, i.e., the counting rate. We determined that correction for sensitivity normalization in variation of activity can be performed using only the triple-coincidence rate as an index, even if using a different nuclide from that used for normalization. From analysis of dual-tracer phantom images using $^{18}$F and $^{44m}$Sc or $^{18}$F and $^{22}$Na, data subtraction in the sinogram data with sensitivity correction gives a higher quality of isolated images for the pure positron emitter than those from image subtractions. Furthermore, from dual-isotope ($^{18}$F-FDG and $^{44m}$Sc) mouse imaging, we concluded that our developed method can be used for practical imaging of a living organism.

[1]Corresponding author.

# Contents

## 1. Introduction

Positron emission tomography (PET) is a powerful tool for radio-tracer imaging in a living biological object. PET was mainly developed for nuclear medicine imaging, such as clinical diagnoses and preclinical studies [1]. Contemporary PET systems consist of a large number of finely pixelized detectors and fast signal processing devices [2-4]. In addition to the development of PET devices, image reconstruction methods have been also developed to obtain high spatial resolution and quantitative images from acquired PET data. Image reconstruction methods have developed from primary simple projection methods to iterative calculation methods as computational power has increased over time[5].

    Consequently, recent PET gives high quality images for various tracers and can be used widely, from basic science to clinical diagnosis. Unfortunately, conventional PET is useful only for single tracer imaging due to the energy constancy of annihilation photons, which are used for PET imaging. In order to improve PET imaging, we have developed a new PET system that can be used for multiple-tracer simultaneous imaging. Our PET system, named multiple-isotope PET (MI-PET), detects not only annihilation photons but also prompt γ-rays, which are emitted successively after positrons, using additional γ-ray detectors. Since a prompt γ-ray has an energy intrinsic to each radio-isotope, MI-PET can identify the tracer radio-isotopes by using prompt γ-ray detection. Previously, we succeeded in proving the basic principle of MI-PET using a prototype system [6]. However, because most image reconstruction methods for PET

imaging presuppose single tracer imaging, study of image reconstruction methods specifically for multiple-tracer imaging is necessary along side MI-PET scanner development.

In previous research, several PET systems dedicated to multiple-tracer imaging have been developed by a few groups [6-10], and a small number of specific image reconstruction methods for multiple-tracer PET imaging have been reported. One method based on simulational data was used by Andreyev et al., in which they simulated a normal PET scanner in the case of a prompt γ-ray detected by a PET detector, and they proposed a specific image reconstruction method based using a random distribution to compensate for different data amounts between the normal PET events and the prompt γ-ray coincidence events [7].

By contrast, our MI-PET system is special in terms of using additional detectors for prompt γ-ray detection, and therefore we need to develop an image reconstruction method specific for our system. In particular, because of the imperfectness of prompt γ-ray detection in MI-PET imaging with a pure positron emitter and positron-γ emitter an image for a pure positron emitter taken by MI-PET is superposed by a positron-γ emitter. Therefore, in order to make an isolated image of the pure positron emitter, we developed a method to make an isolated image of the pure positron emitter and also a system-specific dead-time correction method. We conducted dual-isotope phantom and mouse imaging, and reconstructed images that used the developed methods were evaluated from both quantitative and practical qualitative viewpoints.

## 2. Multiple-isotope PET

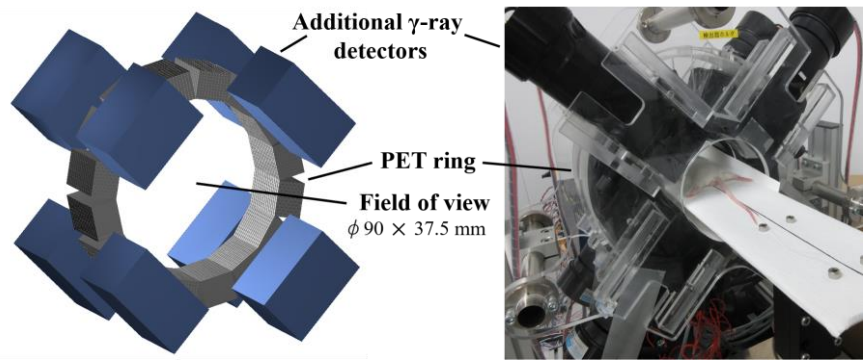

**Figure 1.** Schematic illustration (left) and photograph (right) of MI-PET.

MI-PET consists of a ring-type small-animal PET system and additional γ-ray detectors. The PET scanner is fabricated using pixelized gadolinium orthosilicate scintillators. Eight large volume ($50 \times 50 \times 30$ mm$^3$) bismuth germanium oxide (BGO) detectors are used for the detection of prompt γ-rays, and they are arranged into two rings mounted on each side of the PET ring. The inner diameter of the PET ring is 95 mm, and axial length of the field-of-view (FOV) is 37.5 mm. A schematic illustration and a photography of MI-PET are shown in figure 1.

The signal processing system of MI-PET can acquire list-mode data of both coincident annihilation photons and annihilation photons with coincident prompt γ-rays in parallel. The width of the coincidence time windows is set to 20 ns. Events of coincident annihilation photons are used for conventional PET imaging, and events of coincident annihilation photons with the presence or absence of prompt γ-rays are used for tracer identification.

For our purposes, coincidence detection of annihilation photons by PET detectors is denoted as double-coincidence, and coincidence detection of annihilation photons with prompt γ-ray coincidence by an additional detector is denoted as triple-coincidence.

The list-mode data are classified into two groups based on the presence or absence of prompt γ-rays. Different images of tracers are obtained by using a conventional image reconstruction method. Details of the system and its basic performance have been described by Fukuchi et al. [6].

The sensitivity for a prompt γ-ray of 1275 keV ($^{22}$Na prompt γ-ray) in the center position is about 7%. This sensitivity includes not only full-energy absorption events but also Compton partial energy deposited events of the energy threshold setting in 1000 keV. The energy resolution of an additional BGO detector is about 10%. The energy threshold of a BGO detector is changed according to the nuclide and set at 1000 keV for $^{22}$Na or 800 keV for $^{44m}$Sc (1157-keV prompt γ-ray).

## 3. Methods

The simplest MI-PET operation for dual-isotope imaging uses a pure positron-emitter and a positron-γ emitter. In this case, a reconstructed image from an event data set with the presence of prompt γ-ray detection (triple-coincidence) reflects the distribution of the positron-γ emitter. Because of the imperfection of prompt γ-ray detection, the reconstructed image from the absence of prompt γ-ray detection (double-coincidence) contains a distribution of both the pure positron emitter and the positron-γ emitter. Therefore, a method to reconstruct the isolated image of the pure positron emitter is required, and that method and its performance evaluation are described in this paper.

For our purposes, tracer nuclides used as pure positron emitters and those used as positron γ-ray emitters are called tracer-A and tracer-B, respectively, and the isolated images for tracer-A and -B are called image-A and image-B, respectively. The images reconstructed from the data sets without and with prompt γ-ray detection are called image-D (double-coincidence) and -T (triple-coincidence), respectively.

In both image-D and -T, the positional dependences of scanner sensitivities were normalized by normalization scan data using a cylindrical phantom with a pure positron emitter, $^{18}$F. The position dependences of the prompt γ-ray detection sensitivities were also normalized using a cylindrical phantom with positron-γ emitters $^{44m}$Sc and $^{22}$Na. The details of the normalization scans and their results are described in Sections 4.1 and 5.1, respectively.

In this section, we describe isolation methods for image-A. All isolation methods are based on subtraction between reconstructed image or sinogram data for image-D and image-T. We also describe a correction method for subtraction parameters depending on the activities of measured objects.

### 3.1 Image subtraction

A primitive method to obtain an isolated image of tracer-A is given by subtraction between reconstructed image-D and image-T. Each component in voxel $i$ of image-D ($D_i$) and -T ($T_i$) is expressed using the prompt γ-ray detection efficiency for each voxel position $\varepsilon_{pi}$ as

$$D_i = A_i + B_i \qquad (3.1)$$
$$T_i = \varepsilon_{pi} B_i, \qquad (3.2)$$

where $A_i$ and $B_i$ are the contents of actual activities in voxel $i$ for tracer-A and -B, respectively. These operations are applied for each voxel, but hereafter voxel index $i$ is omitted. From eq. (3.2), $B$ is represented by

$$B = \frac{1}{\varepsilon_p} T, \qquad (3.3)$$

therefore, from eq. (3.1), isolated image $A$ is represented as

$$A = D - \frac{1}{\varepsilon_p} T. \qquad (3.4)$$

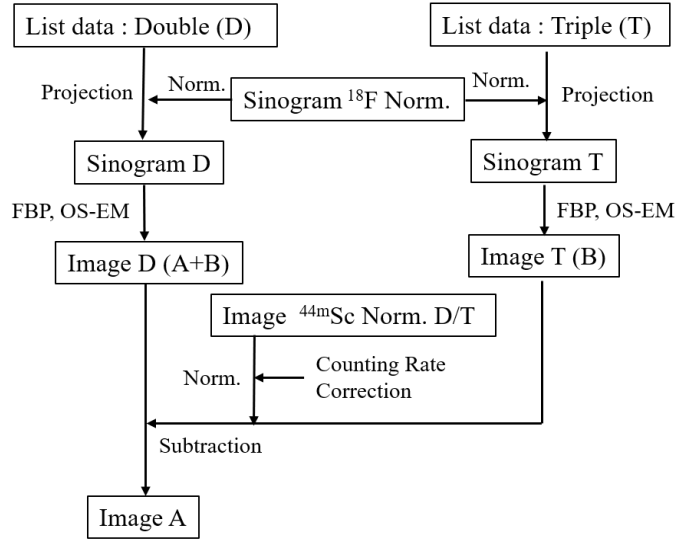

**Figure 2.** Data flow diagram to reconstruct isolated image of tracer-A using image subtraction.

Due to the finite accuracy of measurement, a reconstructed image is different from a true tracer distribution, therefore images $A$ and $B$ contain image noise, artifacts and etc. The detection efficiency of the triple-coincidence $\varepsilon_t$ in each voxel position is represented as

$$\varepsilon_t = \varepsilon_d \varepsilon_p, \qquad (3.5)$$

where $\varepsilon_d$ is the detection efficiency for the double-coincidence. These double- and triple-coincidence efficiencies in each voxel position are determined by a normalization scan using tracer-B. From eq. (3.5), prompt γ-ray efficiency $\varepsilon_p$ is written as

$$\varepsilon_p = \frac{\varepsilon_t}{\varepsilon_d}. \qquad (3.6)$$

Then, from eq. (3.4), image A is

$$A = D - \frac{\varepsilon_d}{\varepsilon_t} T. \qquad (3.7)$$

The whole procedure to reconstruct an isolated image of tracer-A from list mode data is shown in figure 2.

### 3.2 Sinogram subtraction

The other method to reconstruct an isolated image for tracer-A is subtraction using sinogram projection data. In this method, the subtraction coefficient is also determined by the ratio of triple- to double-coincidence by the normalization scan using tracer-B. In the same manner as the image subtraction method (eq. (3.7)), an isolated sinogram for tracer-A ($P_A$) is obtained by

$$P_A = P_D - \frac{\varepsilon'_t}{\varepsilon'_d} P_T, \tag{3.8}$$

where $P_D$ and $P_T$ are sinograms from double- and triple-coincidence data, and $\varepsilon'_d$ and $\varepsilon'_t$ are detection efficiencies for sinograms of double- and triple-coincidence data derived by the normalization scan, respectively. As with image subtraction, voxel index $i$ is omitted, but this operation is applied for each pixel of the sinogram. From isolated sinogram $P_A$, image-A is reconstructed by using a conventional image reconstruction method. This procedure is summarized in figure 3.

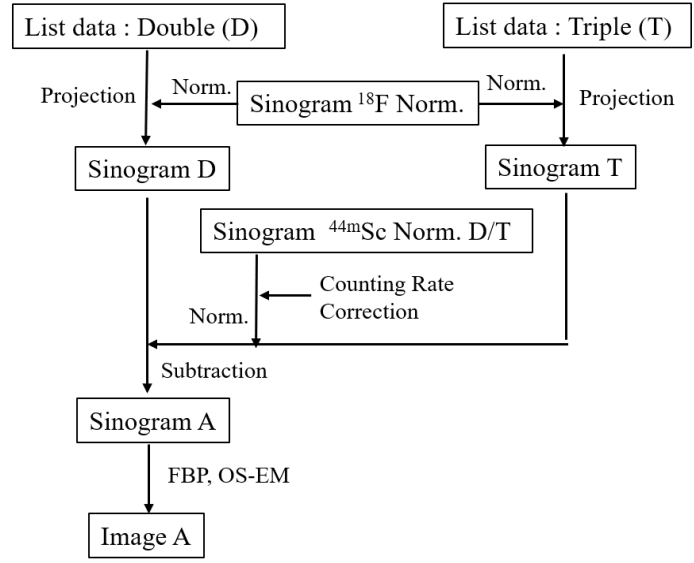

**Figure 3.** Data flow diagram to reconstruct isolated image of tracer-A using sinogram subtraction.

### 3.3 Counting rate correction

Due to the system dead-time difference of single-event acquisition between double- and triple-coincidence, the data acquisition ratio of double- and triple-coincidence events changes depending on the tracer activities in our system. To correct this variation in the reconstruction of isolation images of tracer-A, we developed a method to correct the subtraction coefficient depending on the counting rate.

In our system, the data acquisition system, which causes the main component of dead-time, works with a non-paralyzable behaviour. In general, the recorded count rate $m$ with a non-paralyzable model is

$$m = \frac{n}{n\tau + 1}, \tag{3.9}$$

where $n$ and $\tau$ are the true detection rate and the system dead-time for single-event recording, respectively [11]. Applying this equation for the recorded counting rate of double- ($m_2$) and triple- ($m_3$) coincidences,

$$m_2 = \frac{n_2}{n_2 \tau_2 + 1}, \text{ and} \tag{3.10}$$

$$m_3 = \frac{n_3}{n_3 \tau_3 + 1}, \tag{3.11}$$

where $n_2$ and $n_3$ are the true detection rates for double- and triple-coincidence, respectively, and $\tau_2$ and $\tau_3$ are the dead-times for single-event recording in double- and triple-coincidence, respectively.

The true counting rates can be expressed using detection efficiencies and activity $A$ Bq,

$$n_2 = \varepsilon_2 A \quad \text{(double-coincidence), and} \quad (3.12)$$

$$n_3 = \varepsilon_3 A \quad \text{(triple-coincidence),} \quad (3.13)$$

where $\varepsilon_2$ and $\varepsilon_3$ are the efficiencies for double- and triple-coincidence, respectively. Using these equations, recorded counting rates are

$$m_2 = \frac{\varepsilon_2 A}{\varepsilon_2 A \tau_2 + 1}, \text{ and} \quad (3.14)$$

$$m_3 = \frac{\varepsilon_3 A}{\varepsilon_3 A \tau_3 + 1}. \quad (3.15)$$

From eqs. (3.12) and (3.13), the true counting rate for triple-coincidence is

$$n_2 = \frac{\varepsilon_2}{\varepsilon_3} n_3. \quad (3.16)$$

Accordingly, the ratio of recorded count rates between triple- and double-coincidence $f$ is

$$f = \frac{m_3}{m_2} = \frac{\frac{\varepsilon_2}{\varepsilon_3} n_3 \tau_2 + 1}{\frac{\varepsilon_2}{\varepsilon_3}(n_3 \tau_3 + 1)}. \quad (3.17)$$

Using the efficiency ratio between double- and triple-coincidence ($\varepsilon_2/\varepsilon_3$) from the normalization scan of tracer-B, the single-event dead-time of $\tau_2$ and $\tau_3$ can be determined by fitting eq. (3.17) as a function of triple-coincidence count rate $n_3$.

The image subtraction (eq. (3.7)) to reconstruct an isolated image for tracer-A is corrected by the variation of ratio $m_3/m_2$ as a function of $n_3$ using

$$A = D - \frac{f_N}{f_I} \frac{\varepsilon_d}{\varepsilon_t} T, \quad (3.18)$$

where $f_N$ and $f_I$ are count rates between triple- and double-coincidences (eq. (3.17)) in normalization and imaging scan, respectively. The same correction is applied for sinogram subtraction (eq. (3.8)) using correction parameter.

Using activity decay of positron-γ emitter $^{44m}$Sc, which has a half-life of 58.6 hours, recorded counting-rates for double- and triple-coincidence in the normalization scan were determined and are described in Sections 4.1 and 5.1. Furthermore, because eq. (3.17) depends only on the triple-coincidence detection rate, in the case using tracer-B other than $^{44m}$Sc, which has different prompt γ-ray energy, counting rate correction can be applied based on a $^{44m}$Sc normalization scan.

## 4. Measurements

### 4.1 Normalization scans

For the spatial normalizations of the PET and the prompt γ-ray sensitivity, long-period measurements were performed using cylindrical phantoms of $^{18}$F, $^{44m}$Sc, and $^{22}$Na. Each cylindrical phantom was 180 mm in length and 78 mm in diameter. The volume of the cylindrical phantom inside of the FOV was 20.8% of the total volume. Fluorine-18 and $^{22}$Na were dissolved in water and $^{44m}$Sc was dissolved in HCl (0.5 M) to pour into the phantom vessels.

The normalization scan of pure positron emitter $^{18}$F was used for sensitivity normalization of the PET detectors. The initial activity of $^{18}$F measurement was 63.1 MBq, and the measuring time was 12 hours. Because the half-life of $^{18}$F is 109.8 min, almost all activity decayed during measurement.

The normalization scans of positron-γ emitters $^{44m}$Sc and $^{22}$Na were performed for the normalization of the prompt γ-ray sensitivities. $^{44m}$Sc was also used for analysis of the counting rate dependence using its proper decay half-life of 58.6 hours. The initial activity of the $^{44m}$Sc phantom was 2.45 MBq, and the measuring time was 235 hours, which corresponds to about 4 half-lives.

Scandium-44m was produced *via* the reactions of $^{45}$Sc(*d*, p2n)$^{44m}$Sc (for phantom imaging) and $^{44}$Ca(*d*, 2n)$^{44m}$Sc (for mouse imaging) with a 24-MeV deuterium beam and was purified by chemical processes at the RIKEN AVF cyclotron. Since there were about 48 hours of purification and transportation time from the production site to our facility, a short-lived $^{44}$Sc ($T_{1/2}$ = 6.84 hours) and by-products were attenuated.

The $^{22}$Na cylindrical phantom had 1.1 MBq of activity, and the scanning time was 87 hours.

### 4.2 $^{18}$F + $^{44m}$Sc rod phantom imaging

In order to evaluate the quantitative performance of our image reconstruction methods, we conducted two phantom experiments with dual isotopes (experiment 1: $^{18}$F and $^{44m}$Sc, experiment 2: $^{18}$F and $^{22}$Na). These phantoms are referred to as the F-Sc phantom (experiment 1) and the F-Na phantom (experiment 2).

**Table 1.** Nuclides and activities in rod phantom experiments

|       | Exp. 1 | Exp. 2 |
|-------|--------|--------|
| Rod 1 | $^{18}$F 546 kBq | $^{18}$F 636 kBq |
|       | $^{44m}$Sc 497 kBq | $^{22}$Na 585 kBq |
| Rod 2 | $^{18}$F 546 kBq | $^{18}$F 636 kBq |
| Rod 3 | $^{44m}$Sc 497 kBq | $^{22}$Na 585 kBq |

In the first experiment, $^{18}$F and $^{44m}$Sc were used as tracer-A and -B, respectively. Three rods were arranged in an equilateral triangle at a distance of twice the rod diameter. The configuration of the rod phantom is shown in figure 4. The rods were positioned at the scanner's center parallel to the scanner's axial direction. Fluorine-18 and $^{44m}$Sc dissolved in 1 M HCl were poured into the rods. The first rod had $^{18}$F (546 kBq at the start of the measurement) and

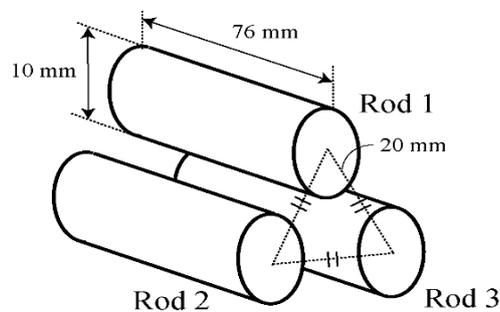

**Figure 4.** Illustration of dual-tracer rod phantom configuration for F-Sc phantom and F-Na phantom.

$^{44m}$Sc (497 kBq) activities. The second and third rods had $^{18}$F (546 kBq at the start of the measurement) and $^{44m}$Sc (497 kBq) activities, respectively. These activities are summarized in table 1. A 30-min scan was performed. The energy threshold of the BGO detectors was set to 800 keV to match the 1157-keV prompt γ-ray of $^{44m}$Sc.

### 4.3 $^{18}$F + $^{22}$Na rod phantom imaging

To evaluate the performance of our method for a nuclide with a different energy from the prompt γ-ray from $^{44m}$Sc, phantom data using $^{18}$F and positron-γ emitter $^{22}$Na were used. This experimental data has been previously reported [6].

The phantom configuration was the same as that of the F-Sc phantom (figure 4). The first rod had $^{18}$F (686 kBq at the start of the measurement) and $^{22}$Na (585 kBq) activities. The second and third rods had $^{18}$F (686 kBq at the start of the measurement) and $^{22}$Na (585 kBq) activities, respectively. These nuclides and activities are summarized in table 1. The energy threshold of the BGO detectors was set to 1000 keV for the prompt γ-ray of $^{22}$Na. The measurement duration was 30 min.

### 4.4 $^{44m}$Sc + $^{18}$F-FDG mouse scan

To evaluate the practical performance of our methods, mouse scanning with a dual-isotope was performed using $^{18}$F-FDG and a simple-substance of $^{44m}$Sc. In this experiment, 1.13-MBq $^{18}$F-FDG and 1.23-MBq $^{44m}$Sc were administered to an 8-week-old normal male mouse by a tail vein injection. After 38 min from administration, a 30-min scan with bed motion was performed under anaesthesia.

This animal experiment was performed in accordance with the Principles of Laboratory Animal Care (NIH Publication No. 85-23, revised 1985) and approved by the Institutional Animal Care and Use Committee (IACUC) of RIKEN, Kobe Branch.

## 5. Results

For all experiments, the acquired list-mode data were categorized into two groups by the presence or absence of prompt γ-ray detection. Two kinds of sinogram data were made from each categorized group of data. The image reconstruction from the sinogram data was performed using an inter-update Metz-filtered ordered subset expectation maximization by a program called Software for Tomographic Image Reconstruction [12, 13] with four iterations and four subsets. To eliminate random coincidence background, delayed coincidence components were subtracted from true coincidence components. Because the phantoms and animals used for measurements had small volumes, photon absorptions in the imaged objects were ignored. In this section, we describe the results of analysis for the isolated images of tracer-A using various methods.

### 5.1 Normalization scans

The normalization scan data of $^{18}$F was used for sensitivity normalization of the PET detectors. The sinogram projection data was applied for the correction of each imaging projection data group. Using normalization scan data of $^{44m}$Sc and $^{22}$Na, the position dependence of the sensitivity for the prompt γ-ray, i.e., the subtraction parameters for making an isolated image of tracer-A, was determined for each nuclide.

In addition to sensitivity normalization, $^{44m}$Sc normalization data was used for evaluation of the counting rate dependence of the prompt γ-ray sensitivity as a ratio of double- and triple-coincidence. The counting rate of the double- and triple-coincidence as a function of measurement time in $^{44m}$Sc is shown in figure 5 (i). In this graph, the counting rate for triple-coincidence is ten-fold that of double-coincidence. The activity of $^{44m}$Sc in the FOV at the start was 510 kBq, and the counting rate for triple-coincidence at that time was 136.9 cps. Since the measuring time of 235 hours corresponds to 4 times the half-life of $^{44m}$Sc (58.6 hours), the counting rates of both double- and triple-coincidences were decreased according to activity decay. The curves fitted by the exponential functions (solid line) and mathematical expressions are also shown in figure 5 (i). From these fittings, the attenuation coefficients were different between double- and triple-coincidence. This is because of the difference of the dead-time for the single event processing time for double- and triple-coincidences, as mentioned in Section 3.3.

The functions obtained by division of triple- and double-coincidence as a function of the triple-coincidence counting rate are shown in figure 5 (ii). The variation of this value indicates the necessity of correction in data subtraction for the reconstruction of isolation images of tracer-A. The ratio from a total event of a normalization scan of $^{44m}$Sc was determined to be 57.0 cps, and this value correspond to the time point at $2.69 \times 10^3$ sec. This value was indicated in figure 5 (ii). The correction factor $f$ in eq. (3.18) for imaging using the $^{44m}$Sc tracer is determined by the difference between this average value and the counting rate in the imaging.

The triple-coincidence counting rate of the $^{22}$Na normalization scan is constant during measurement because of the 2.6-year half-life, and the counting rate was determined to be 86.7 cps (shown in figure 5 (ii)). For the imaging with $^{22}$Na, which has a different energy from the prompt γ-ray from $^{44m}$Sc, assuming the correction factor depended only on the counting rate even if the efficiency for the prompt γ-ray is different, the subtraction parameter $f$ for $^{22}$Na imaging was deduced using the ratio between the counting rates in the $^{22}$Na normalization measurement and the imaging object using the function in figure 5 (ii).

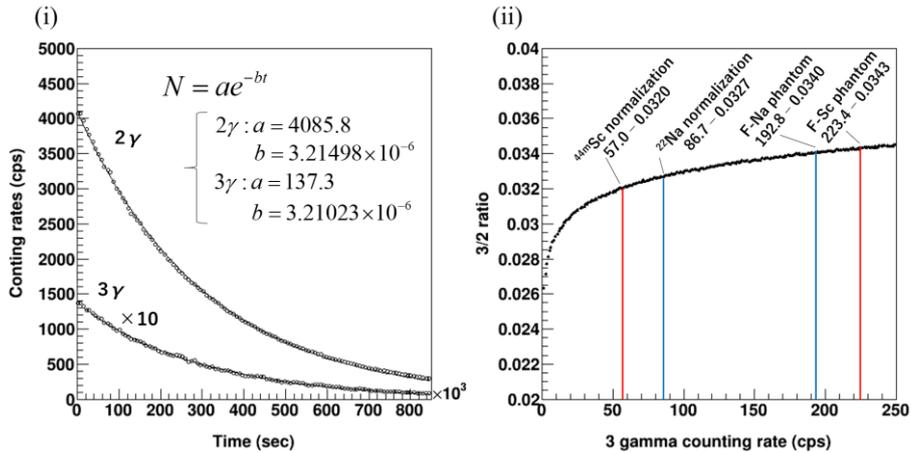

**Figure 5.** Results of normalization scan with $^{44m}$Sc. (i) Counting rate of double- and triple-coincidence as a function of measurment time in $^{44m}$Sc (The counting rate for triple-coincidence is ten-fold that of double-coincidence). (ii) Functions obtained by division of triple- and double-coincidence as a function of triple-coincidence counting rate.

## 5.2 $^{18}$F + $^{44m}$Sc rod phantom

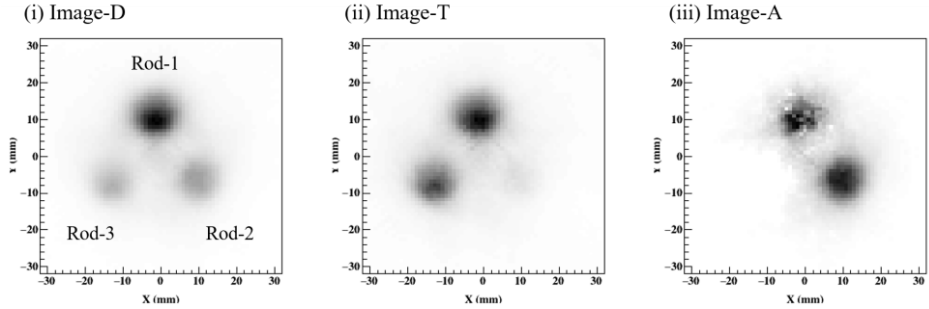

**Figure 6.** F-Sc phantom images in xy-projection. (i) Image reconstructed from double-coincidence data (image-D, $^{18}$F and $^{44m}$Sc). (ii) Image reconstructed from triple-coincidence data (image-T, $^{44m}$Sc). (iii) Isolated image of tracer-A ($^{18}$F) by subtraction between image-D and image-T.

The reconstructed images in xy-projection for the F-Sc phantom with the absence (image-D) and presence (image-T) of prompt γ-ray detection are shown in figure 6 (i) and (ii), respectively. Distributions of both tracer-A and -B appear in image-D, and the distribution of tracer-B appears in image-T. An isolated image of tracer-A obtained by image subtraction between image-D and image-T (method in Sec. 3.1) is shown in figure 6 (iii). In this subtraction, sensitivity normalization in each voxel using eq. (3.7) was performed using $^{44m}$Sc normalization of the total data.

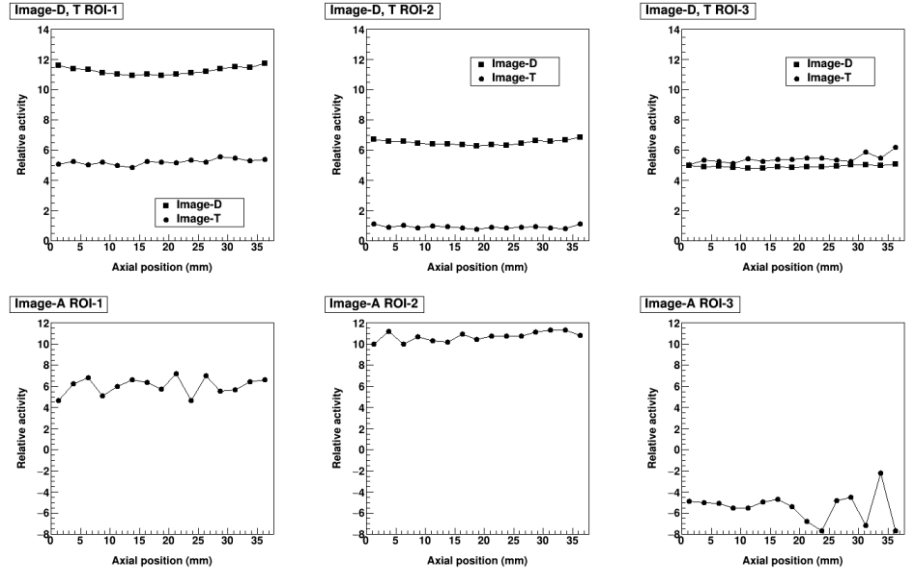

**Figure 7.** Results of ROI analysis for F-Sc phantom images. Upper row: results of image-D and image-T. Lower row: isolated image-A.

To analyze quantitatively these images, regions of interest (ROIs) were set on each rod of the xy-slices, and values in the ROIs are shown in figure 7. The upper row shows values for image-D and image-T, and the lower row shows them for isolated image-A. The ROI numbers correspond to the rod numbers. In the isolated image of tracer-A, ROI-2 values are over-subtracted, and the image for tracer-A (figure 6 (iii)) has large statistical fluctuations. In addition, ROI analysis for this image shows worse uniformity than for image-D and image-T.

Isolated image-A with counting rate corrections (method in Section 3.3, eq. (3.18)) reconstructed by image subtraction using eq. (3.7) and sinogram data subtraction using eq. (3.8) are shown in figure 8 (i) and (ii), respectively. Correction factor $f_N/f_I$ was common for both images and was determined to be 0.933 from the triple-coincidence counting rate on F-Sc phantom imaging of 223.4 cps (as indicated in figure 5 (ii)). Results of the ROI analyses for these images (upper row: image subtraction; lower row: sinogram subtraction) are shown in figure 9. Over-subtraction is resolved in the reconstructed image of subtraction with rate correction (figure 9 (i)), however, statistical fluctuations still remain. While over subtraction is also resolved in the reconstructed image by sinogram data subtraction (figure 9 (ii)), image uniformity in the xy-slices is comparable with that in image-D and image-T.

In conclusion, the isolated image of tracer-A from the sinogram data subtraction with counting rate correction gives the best quality. Therefore, hereafter, isolated images of tracer-A in the Na-F phantom and mouse experiments were reconstructed by using this method.

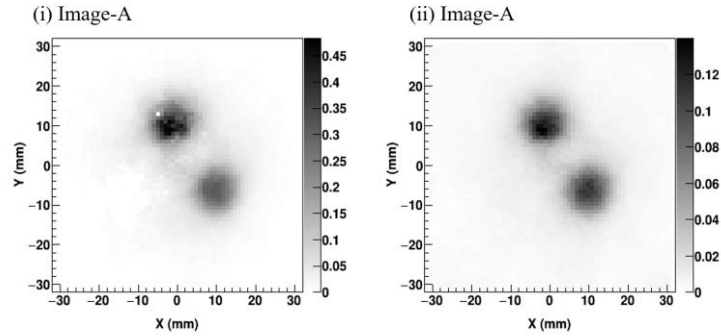

**Figure 8.** Isolated image-A ($^{18}$F) with couting rate correction. (i) Isolated image-A by subtraction of images. (ii) Isolated image-A by subtraction of sinograms.

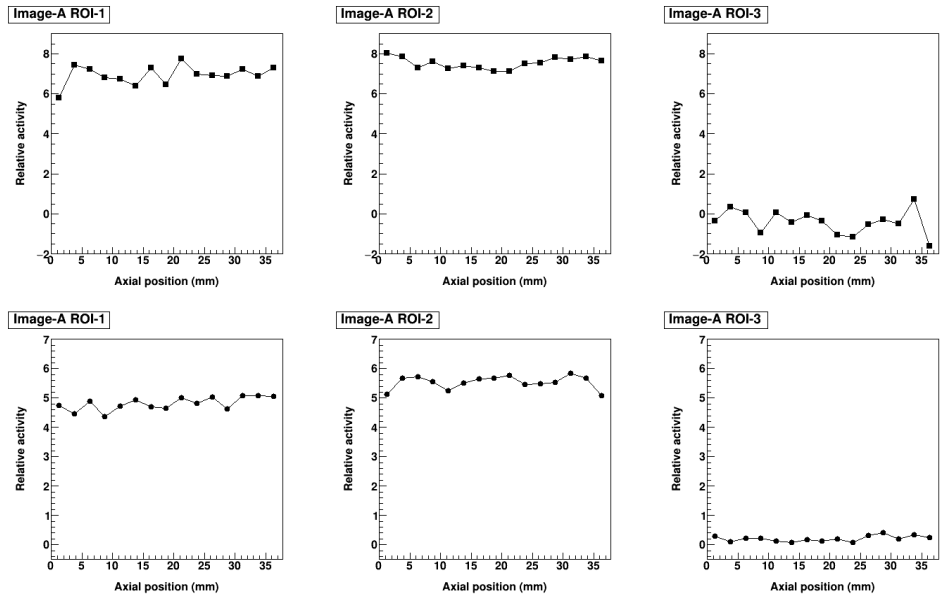

**Figure 9.** Results of ROI analysis for isolated image-A with counting rate corrections. (i) Isolated image-A by subtraction of images. (ii) Isolated image-A by subtraction of sinograms.

## 5.3 $^{18}$F + $^{22}$Na rod phantom

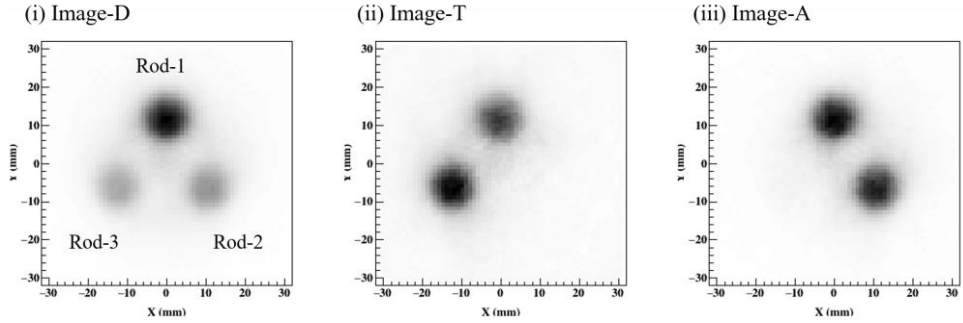

**Figure 10.** F-Na phantom images in xy-projection. (i) Image reconstructed from double-coincidence data (image-D, $^{18}$F and $^{22}$Na). (ii) Image reconstructed from triple-coincidence data (image-T, $^{22}$Na). (iii) Isolated image of tracer-A ($^{18}$F) by subtraction between sinogramas with counting rate correction.

The reconstructed images in xy-projection for the F-Na phantom with the absence (image-D) and presence (image-T) of prompt γ-ray detection are shown in figure 10 (i) and (ii), respectively. The isolated image of tracer-A constructed by subtraction of the sinogram data with counting rate correction is shown in figure 10 (iii).

To determine the prompt γ-ray efficiency as a subtraction parameter, $^{22}$Na normalization scan data was used. Counting rate correction factor was determined by extrapolation of the curve in figure 5 (ii) as the triple-coincidence counting rate from $^{22}$Na normalization (86.2 cps) to this phantom (192.8 cps). Correction factor $f_N/f_I$ was determined to be 0.962.

ROI analysis in the same manner as with F-Sc phantom was performed, and those results are shown in figure 11. From projection images and ROI analyses, image-A reconstructed by sinogram subtraction with counting rate correction had equivalent qualities with image-D and image-T.

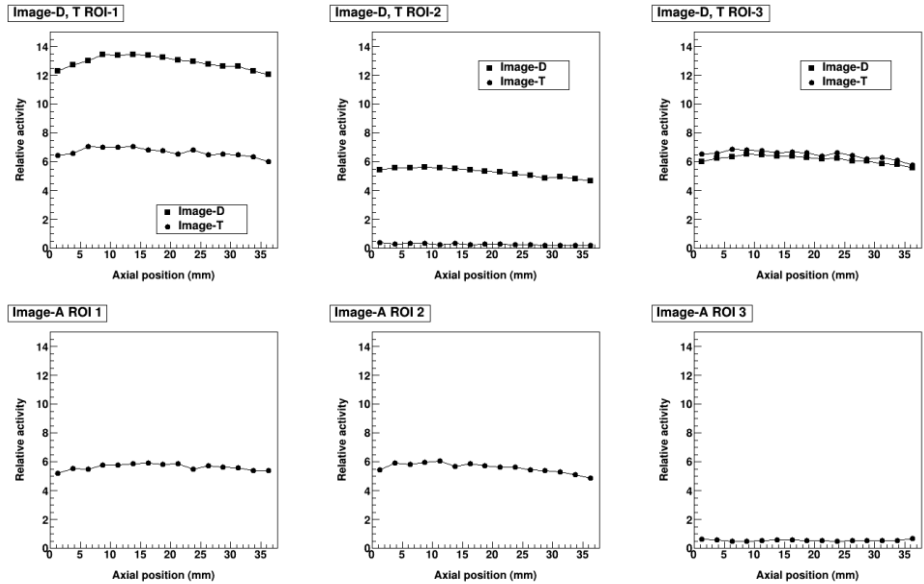

**Figure 11.** Results of ROI analysis for F-Na phantom images. Upper row: results of image-D and image-T. Lower row: isolated image-A.

### 5.4 $^{18}$F-FDG + $^{44m}$Sc mouse

Similar to the rod phantoms, an isolated image of tracer-A for the mouse scan was reconstructed using a subtraction between sinogram data with absence and presence of the prompt γ-ray detection. The subtraction parameter was determined by using $^{44m}$Sc normalization scan data, and counting rate correction was also performed based on the counting rate in each bed position.

Reconstructed image-D, image-T, and isolated image-A are shown in figure 12 (i), (ii), and (iii), respectively. From image-T, we can clearly observe $^{44m}$Sc accumulation in the liver, whereas from isolated image-A, $^{18}$F-FDG accumulated in the heart and urinary bladder. These distributions are reasonable from a physiological viewpoint.

To make a quantitative comparison, 3D volumes of interest (VOIs) were set on the heart, liver, and bladder for image-D, image-T, and image-A. Result values of these VOIs are shown in table 2. The activity values in table 1 are indicated by the positron emission rate, i.e., normalized by the positron emission rates ($^{18}$F: 96.73%, $^{44m}$Sc: 94.27%). For the activity of $^{18}$F-FDG, considering attenuation during the scan, the accumulated activities were calculated using an average value during the scan. Direct subtraction of VOI values between image-D and image-T (D-T) are also shown in table 2.

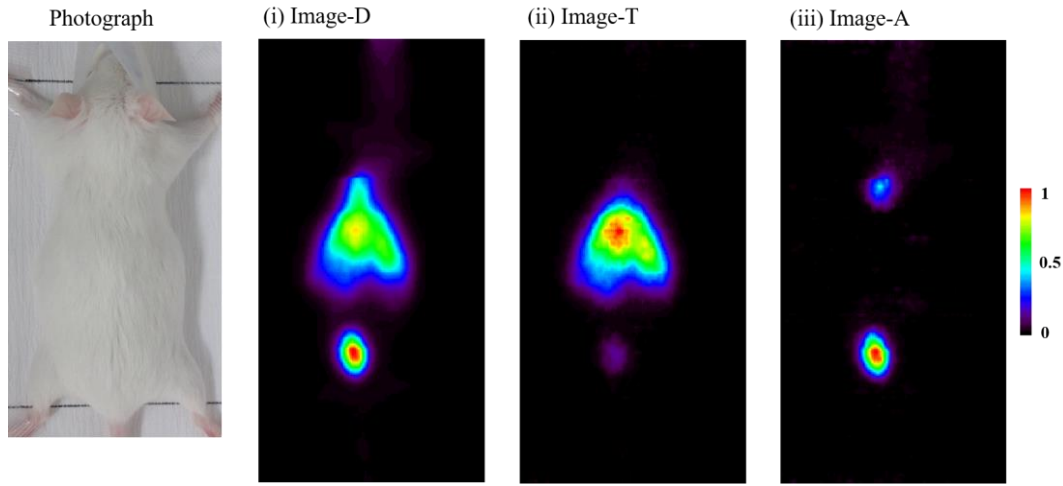

**Figure 12.** Photograph (left) and reconstructed images for $^{18}$F-FDG and $^{44m}$Sc administrated mouse. (i) image-D ($^{18}$F and $^{44m}$Sc), (ii) image-T ($^{44m}$Sc), (iii) image-A ($^{18}$F) reconstructed by sinogram subtraction with counting rate correction.

**Table 2.** ROI analysis of mouse. ROI values are shown in kBq.

|         | Image-D | Image-T | D-T  | Image-A |
|---------|---------|---------|------|---------|
| Heart   | 31.7    | 10.2    | 21.5 | 23.9    |
| Liver   | 421.3   | 427.6   | -6.3 | 0.9     |
| Bladder | 69.0    | 9.2     | 59.8 | 77.3    |

## 6. Discussion

From the analysis of the F-Sc phantom data, to reconstruct an isolated image of tracer-A by data subtraction, data subtraction in sinogram data using $^{44m}$Sc normalization data gave better quality image data than did subtraction in reconstructed image data. A reason that sinogram subtraction gives a better quality isolated image-A than image subtraction is probably the less biased distribution of sinograms differed from images that have partial volume effects or other effects. Furthermore, if measured activity is different from normalization scanning, correction for the ratio between double- and triple-coincidence is needed in image subtraction to reconstruct an isolated image of tracer-A. For this correction, only the counting rate of triple-coincidence is an indicator for this activity correction.

From the F-Na phantom imaging, for the counting rate correction for $^{22}$Na, which emits prompt γ-rays with different energy and gives different detection efficiency and position dependence of the prompt γ-rays, we found that the normalization data of $^{22}$Na can be scaled only by the triple-coincidence rates. This correction is useful for a long-lived nuclide, such as $^{22}$Na, for which normalization scanning with activity function is difficult. However, this method is also useful for a short-lived nuclide, in which it can be difficult to acquire a sufficient amount of normalization data.

From the VOI analysis of the mouse experiment with $^{18}$F-FDG and $^{44m}$Sc, the D-T activity of $^{18}$F-FDG (table 2) in the liver was negative. However, activity in the liver of image-A, which has no biological specific accumulation of $^{18}$F-FDG, was nearly zero and was comparable with other non-accumulated sections. From this result, we determined that our newly developed image-A isolation and counting rate correction method are useful for practical dual-isotope imaging using MI-PET.

Our method can probably be applied not only to MI-PET imaging using dual-isotope tracer-A and tracer-B, but also to those such as tracer-B and tracer-B', or for three tracers with tracer-A, tracer-B, and tracer-B'.

Furthermore, our correction method can be also applied for similar types of systems, namely parallel data acquisition for different types of events, which have different data-times.

## 7. Conclusion

To make an isolated image of tracer-A, we developed an image reconstruction method based on data subtraction specific for MI-PET. For prompt γ-ray sensitivity normalization that is used as a subtraction parameter, we performed normalization scans using cylindrical phantoms of the positron-γ emitters $^{44m}$Sc and $^{22}$Na.

From a long-period measurement using the activity decay of $^{44m}$Sc, the ratio between the prompt γ-ray and the PET sensitivity changes as a function of the activities. Therefore, we developed a subtraction parameter correction method that depends on the activities.

From analyses of dual-tracer phantom imaging, we determined that data subtraction in the sinogram data gives a higher quality of isolated images for the pure positron emitter than those reconstructed images from image subtractions. We determined that sensitivity correction for the normalization factor in variation of activity can be performed using only the triple-coincidence rate as an index even with a different nuclide from that used for normalization.

From ROI analyses of dual-tracer phantom imaging, the isolated image of a pure positron emitter reconstructed by our image reconstruction method with sensitivity correction had a quality equal to that of the original images. Furthermore, from dual-isotope mouse imaging, our method was found to provide practical images in imaging for a living organism.

**Acknowledgments**

The authors thank Mr. H. Mashino, who produced the detection system. This work was supported by JSPS KAKENHI Grant Numbers JP15H04770. The $^{44m}$Sc was supplied through Supply Platform of Short-lived Radioisotopes, supported by JSPS Grant-in-Aid for Scientific Research on Innovative Areas, Grant Number 16H06278.

**Conflict of interest disclosure**

The authors have no relevant conflicts of interest to disclose.

**References**

[1] T. Jones and D. Townsend, "History and future technical innovation in positron emission tomography," Jour. Nucl. Med. 4(1), 011013 (2017).

[2] T. Kaneta, M. Ogawa, N. Motomura, H. Iizuka, T. Arisawa, A. Hino-Shishikura, K. Yoshida, and T. Inoue, "Initial evaluation of the Celesteion largebore PET/CT scanner in accordance with the NEMA NU2-2012 standard and the Japanese guideline for oncology FDG PET/CT data acquisition protocol version 2.0," EJNMMI Research

[3] D. F. C. Hsu, E. Ilan, W. T. Peterson, J. Uribe, M. Lubberink, and C. S. Levin, "Studies of a Next-Generation Silicon-Photomultiplier-Based Time-of-Flight PET/CT System," Jour. Nucl. Med. 58, pp. 1511-1518 (2017).

[4] J. van Sluis, J. de Jong, J. Schaar, W. Noordzij, P. van Snich, R. Dierckx, R. Borra, A. Willemsen, and R. Boellaard, "Performance Cahractristics of the Digital Biography Vision PET/CT System, " Jour. Nucl. Med. 60 pp.1031-1036 (2019).

[5] Fei Gao, Kuangyu Shi, Shuo Li, "Computational Method for Molecular Imaging (Lecture Notes in Computational Vision and Biomechanics)," Springer (2016).

[6] T. Fukuchi, T. Okauchi, M. Shigeta, S. Yamamoto, Y. Watanabe, and S. Enomoto, "Positron emission tomography with additional γ-ray detectors for multiple-tracer imaging," Med. Phys. 44(6), pp. 2257-2266 (2017).

[7] A. Andreyev and A. Celler, "Dual-isotope PET using positron-gamma emitters," Phys. Med. Biol. **56**, pp. 4539-4556 (2011).

[8] R. S. Miyaoka, W. C. J. Hunter, A. Andreyev, L. Pierce, T. K. Lewellen, A. Celler, and P. E. Kinahan, "Dual-radioisotope PET data acquisition and analysis," in *Proceedings of the IEEE Nuclear Science Symposium and Medical Imaging Conference 2011* (IEEE, Valencia, Spain, 2011), pp. 3780-3783 (2012).

[9] E. González, P. D. Olcott, M. Bieniosek, and C. S. Levin, "Methods for increasing the sensitivity of

simultaneous multi-isotope positron emission tomography," in *Proceedings of the IEEE Nuclear Science Symposium and Medical Imaging Conference 2011* (IEEE, Valencia, Spain, 2011), pp. 3597-3601 (2012).

[10] A. Andreyev, A. Sitek, A. Celler, "EM reconstruction of dual isotope PET using staggerd injections and prompt gamma positron emitter, " Med. Phys. 41 022501 (2014).

[11] G. F. Knoll, Radiation *Radiation detection and measurements*, John Willey and Sons, Inc., New York 2000.

[12] M. Jacobson, R. Levkopvitz, A. Ben-Tal, K. Thielemans, T. Spinks, D. Belluzzo, E Pagani,
V. Bettinardi, A. Zverovich, M.C. Gilardi, and G. Mitra, "Enhanced 3D PET OSEM reconstruction using inter-update Metz filtering," Phys. Med. Biol. **45(8)**, pp. 2417-2439 (2000).

[13] K. Thielemans, C. Tsoumpas, S. Mustafovic, T. Beisel, P. Aguiar, N. Dikaios, and M. W. Jacobson, "STIR: software for tomographic image reconstruction release 2," Phys. Med. Biol. **57**, pp. 867-883 (2012).
– 17 –